\documentclass[review,12pt]{elsarticle}

\usepackage{amssymb}
\usepackage{amsthm}
\usepackage{amsmath}
\usepackage{mathrsfs}
\usepackage{graphicx}
\usepackage{epstopdf}
\usepackage{float}
\usepackage{caption}
\usepackage{subcaption}
\usepackage{booktabs}
\usepackage{natbib}
\usepackage{bm}
\usepackage{mathrsfs}
\usepackage{soul}
\usepackage{graphicx}
\usepackage{xcolor}
\usepackage{courier} 
\usepackage{listings} 
\usepackage{tabu} 
\usepackage{longtable}
\usepackage{changepage} 
\usepackage[margin=2cm]{geometry}
\biboptions{sort&compress} 
\usepackage{hyperref}
\usepackage{cleveref}

\usepackage{lineno}

\journal{International Journal of Hydrogen Energy}

\makeatletter
\def\@author#1{\g@addto@macro\elsauthors{\normalsize%
    \def\baselinestretch{1}%
    \upshape\authorsep#1\unskip\textsuperscript{%
      \ifx\@fnmark\@empty\else\unskip\sep\@fnmark\let\sep=,\fi
      \ifx\@corref\@empty\else\unskip\sep\@corref\let\sep=,\fi
      }%
    \def\authorsep{\unskip,\space}%
    \global\let\@fnmark\@empty
    \global\let\@corref\@empty  
    \global\let\sep\@empty}%
    \@eadauthor={#1}
}
\makeatother

\Crefname{equation}{Eq.}{Eqs.}
\Crefname{figure}{Fig.}{Figs.}

\begin{document}

\begin{frontmatter}



\title{Virtual failure assessment diagrams for hydrogen transmission pipelines}


\author[OXFORD]{Job Wijnen}

\author[EPRI]{Jonathan Parker}

\author[EPRI]{Michael Gagliano}

\author[OXFORD]{Emilio Mart\'{\i}nez-Pa\~neda\corref{cor1}}
\ead{emilio.martinez-paneda@eng.ox.ac.uk}

\address[OXFORD]{Department of Engineering Science, University of Oxford, Oxford OX1 3PJ, UK}
\address[EPRI]{Electric Power Research Institute, 3420 Hillview Avenue, Palo Alto, CA 94304, USA}

\cortext[cor1]{Corresponding author.}

\begin{abstract}
We combine state-of-the-art thermo-metallurgical welding process modelling with coupled diffusion-elastic-plastic phase field fracture simulations to predict the failure states of hydrogen transport pipelines. This enables quantitatively resolving residual stress states and the role of brittle, hard regions of the weld such as the heat affected zone (HAZ). Failure pressures can be efficiently quantified as a function of asset state (existing defects), materials and weld procedures adopted, and hydrogen purity. Importantly, simulations spanning numerous relevant conditions (defect size and orientations) are used to build \emph{Virtual} Failure Assessment Diagrams (FADs), enabling a straightforward uptake of this mechanistic approach in fitness-for-service assessment. Model predictions are in very good agreement with FAD approaches from the standards but show that the latter are not conservative when resolving the heterogeneous nature of the weld microstructure. Appropriate, \emph{mechanistic} FAD safety factors are established that account for the role of residual stresses and hard, brittle weld regions.\\ 

\end{abstract}
    

\end{frontmatter}




\section{Introduction}
\label{section:introdcution}

A significant challenge preventing the widespread adoption of hydrogen-based technologies stems from the difficulties of transporting and storing hydrogen. Steels are widely used in critical infrastructures, such as pipeline networks. However, when exposed to hydrogen, steels typically display severely degraded material properties, including reduced ductility and reduced fracture toughness \cite{yu2024hydrogen,chen2024hydrogen}. This phenomenon, known as hydrogen embrittlement, results from hydrogen atoms entering and diffusing throughout the material \cite{guedes2020role,cupertino2023hydrogen}.\\

In addition to the development of dedicated hydrogen energy infrastructure, there is growing interest in repurposing parts of the existing gas pipeline network to transport hydrogen \cite{Wang2022,Lipiainen2023,Hoschke2023,Rosa2025}. This approach can significantly accelerate the transition to a hydrogen-based energy economy by substantially reducing costs and implementation times. However, determining safe operating conditions for hydrogen pipelines poses a significant challenge, involving multiple aspects. First, the mechanical performance of pipeline steels in hydrogen-rich environments must be quantified. 
Therefore, a substantial amount of experimental research has been dedicated to investigating the mechanical properties of such steels when exposed to hydrogen \cite{Nanninga2012,Ronevich2016,Jemblie2024,Myhre2024,Chowdhury2025}. Moreover, pipeline failure typically occurs in welded regions \cite{James2016}. During the welding process of a pipeline, the material close to the weld metal (WM), known as the heat affected zone (HAZ), undergoes multiple thermal cycles. This can significantly alter the microstructure and consequently the mechanical properties of the material; e.g., due to the formation of harder phases \cite{Lancaster1999}. Hard HAZs are more brittle and exhibit a higher susceptibility to hydrogen embrittlement. In addition, residual stresses and imperfections can be introduced as a result of the welding process. This makes the WM and HAZ the most susceptible regions to failure in a pipeline network. For this reason, the effect of hydrogen on the properties of HAZ and WM microstructures has been investigated in numerous studies \cite{Chatzidouros2011,Agnani2023,Ronevich2021,Bortot2024,Martin2022,Chowdhury2024b}. Finally, imperfections and defects can compromise the structural integrity of pipeline networks. Such imperfections can be introduced during the manufacturing, welding, or in-service time of pipelines. Therefore, the assessment of defects in pipeline welds and HAZs exposed to hydrogen is crucial to ensure the safe and reliable operation of hydrogen pipelines. However, the complex interplay of the abovementioned factors makes this a complicated task. Among others, it is difficult to adequately characterise residual stresses and the fracture resistance of the HAZ, due to its small size. Moreover, when considering the retrofitting of natural gas pipelines to transport hydrogen, there are vast combinations of pipe and weld materials, weld procedures, H$_2$ conditions (purity and pressure), and pre-existing defects. Physical testing of all of these conditions while resolving the role of residual stresses and (small) hard, brittle weld regions is a remarkable task. For this reason, computational methods have become an essential tool for assessing defects, gaining insights, and guiding design \cite{martinez2021progress}.\\

Recent advances in computational fracture mechanics, combined with significant increases in computer power, have enabled the development of numerical frameworks capable of accurately predicting the fracture behaviour of full-scale components (\emph{Virtual Testing}) \cite{kristensen2020applications}. Particularly critical to this milestone has been the development of phase field fracture methods, which have shown to be capable of tackling complex fracture mechanics problems, handling arbitrary geometries and crack paths, as well as predicting crack initiation, propagation, coalescence, and branching \cite{Miehe2010b,Tanne2018,bourdin2014morphogenesis,boyce2022cracking}. This success has been extended to the prediction of hydrogen-assisted failures, by accounting for hydrogen uptake \cite{hageman2023phase,cupertino2024suitability}, diffusion \cite{Martinez-Paneda2018,isfandbod2021mechanism} and suitable toughness degradation laws \cite{wu2020phase,valverde2022computational}. Phase field-based predictions of hydrogen embrittlement have provided new physical insight and shown to be in good agreement with experiments \cite{Kristensen2020,singh2024hydrogen,zhao2024phase}.\\

Recently, a computational framework based on the phase field fracture paradigm has been developed to assess the structural integrity of welds in hydrogen transport pipelines \cite{Mandal2024,Wijnen2025}. Mandal \textit{et al}.~\cite{Mandal2024} combined a welding process model, capable of predicting residual stress distributions, with a coupled deformation-diffusion-fracture model to predict critical pressures of hydrogen transport pipelines. Wijnen \textit{et al}.~\cite{Wijnen2025} included phase transformations into the welding process to predict heterogeneous property distribution maps in the weld metal and HAZ, which can subsequently be used in microstructurally-informed fracture simulation. Although these advanced models are excellent tools to assess defects in hydrogen pipelines, as they can account for heterogeneous properties in susceptible weld and HAZ regions, conducting detailed fracture mechanics simulations for each defect in a pipeline network can be a computationally intensive and time-consuming task. Therefore, in practice, so-called fitness-for-service (FFS) assessments of pipeline networks are conducted following industrial standards, such as the BS7910 standard \cite{bs7910} or the API 579-1/ASME FFS-1 standard \cite{asme579}. Typical FFS assessments of a flawed structure are based on failure assessment diagrams (FADs). In a FAD, the structural integrity of a structure is evaluated based on its proximity to the two extreme failure modes: brittle crack propagation, governed by linear elastic fracture mechanics (LEFM), and ultimate plastic collapse. In an FAD, the normalized driving forces of both failure mechanisms are used as axes of the FAD. A failure assessment line (FAL) is constructed which represents the failure envelope resulting from the interaction of both driving forces. The FAL demarcates the safe and unsafe operating regimes of the FAD. A flawed structure under a given load can be plotted as an assessment point in the FAD. This point should be located in the safe region of the FAD.\\ 

Numerous studies in the literature can be found in which an FAD is used to investigate defects in pipelines \cite{Pluvinage2006,Ainsworth2016,Moustabchir2016,Kouzoumis2018,Wang2020,Zangeneh2020}. In addition, several studies have extended the use of FADs to hydrogen pipelines. Arroyo \textit{et al}.~\cite{Martinez2019} suggested accounting for the fracture toughness reduction in the calculation of the LEFM driving force, while the plastic collapse driving force remained unchanged. This approach was also adopted by Pluvinage \textit{et al}.~\cite{Pluvinage2021}. Boukortt \textit{et al}.~\cite{Boukortt2018} also took into account the slightly reduced yield stress due to hydrogen when determining the locations of FAD assessment points for a defect in an X52 pipeline steel. Furthermore, Kim \textit{et al}.~\cite{Kim2024} studied the effect of temperature on the properties that are used to construct FADs for a hydrogen-embrittled material. These material properties were predicted using a damage model that accounts for temperature and hydrogen. Their results showed that temperature had only a minor influence on the positions of assessment points in the FAD. Although FFS assessments based on FADs are widely adopted, there are inherent limitations. Multiple simplifications and assumptions have to be made in the construction of an FAD. Specifically, local hard spots and spatially varying properties in weld and HAZ regions can compromise the reliability of FFS assessments. In other words, existing FAD approaches are only valid to assess the integrity of the base metal and not the pipeline's weakest link. Furthermore, the validity of FADs when components are exposed to hydrogen-rich environments is still an open question.\\

In this study, we combine thermo-metallurgical weld process modelling with phase field-based deformation-diffusion-fracture simulations of weld integrity to study the failure of hydrogen transmission pipelines, building upon the work by Mandal \textit{et al}.~\cite{Mandal2024} and Wijnen \textit{et al}.~\cite{Wijnen2025}. For the first time, this mechanistic framework is used to build \emph{Virtual} Failure Assessment Diagrams (FADs), which account for residual stresses and weld microstructural heterogeneity, the two key unknowns in current procedures. A new class of hydrogen Failure Assessment Lines (FALs) are developed, which resolve detailed property distributions, residual stresses and the role of HAZs, as well as the interplay of these key elements with hydrogen. This allows us to shed new light into the validity and limitations of the assumptions commonly made in the fitness-for-service assessment of hydrogen transmission pipelines, quantifying errors and providing appropriate safety factors. Moreover, the resulting virtual failure envelopes provide a straightforward way of incorporating advanced, detailed simulations into engineering practice.\\

This paper is structured as follows. In \Cref{section:fad}, fundamental aspects of FADs will be discussed. The fracture model will be briefly explained in \Cref{section:model}. In \Cref{section:sent}, a comparison between model predictions and standardised FADs will be performed for a simple single-edge notched tensile (SENT) specimen with homogeneous properties, establishing a connection between the two methods. These results showcase the potential of phase field fracture in delivering virtual FADs in agreement with well-established fracture mechanics theory. Furthermore, the influence of hydrogen on the FAD will be examined. In \Cref{section:pipelines}, a more complex and realistic scenario, involving a range of defects in a pipeline weld exhibiting heterogeneous properties, is considered. Here, the model serves as a reference to assess the validity of FAD assessment under relevant assumptions, providing a better understanding of the limitations and uncertainties.

\section{Failure assessment diagrams (FADs)}
\label{section:fad}

Defect assessment codes provide detailed guidelines on how to construct and use FADs. In this paper, we follow the British Standard BS7910:2019 \cite{bs7910}. The key aspects of FAD assessments are briefly summarized in this section.

\subsection{Principle failure criteria}

An FAD is an interaction diagram based on the two principal failure criteria of a flawed structure, namely, ultimate plastic collapse and linear elastic fracture mechanics (LEFM). Plastic collapse occurs when a substantial part of a structure undergoes significant plastic straining, resulting in unstable deformation. In this situation, failure is not governed by crack propagation.\\

\begin{figure}[!tb]
    \begin{subfigure}{\linewidth}
        \includegraphics[width=\linewidth]{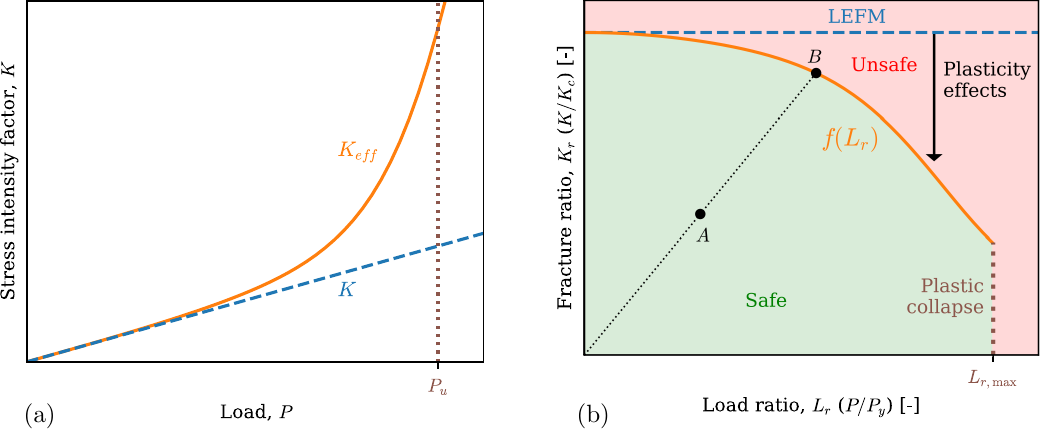}
        \phantomcaption \label{fig:fad_basics_a}
        \phantomcaption \label{fig:fad_basics_b}
    \end{subfigure}
    \caption{Representations of crack driving forces. (a) Elastic stress intensity factor $K$ and effective stress intensity factor as function of applied load. (b) A failure assessment diagram (FAD), indicating safe and unsafe operating regimes.}
    \label{fig:fad_basics}
\end{figure}

Failure due to the propagation of an existing defect in a linear elastic material can be predicted by LEFM. A crack will propagate when the applied stress intensity factor (SIF), $K$, describing the amplitude of the $1/\sqrt{r}$ stress singularity in the vicinity of the crack tip, reaches a critical value $K_c$. For common geometries, SIF solutions are available in handbooks or textbooks. These SIF solutions are of the form
\begin{equation}
    K = \frac{P}{B\sqrt{W}} f\left(\frac{a}{W}\right),
    \label{eq:sif}
\end{equation}
where $P$ is the applied load, $a$ is the crack length, $W$ is the width of the geometry, $B$ is the thickness, and $f(a/W)$ is a geometry-dependent dimensionless function. The SIF, $K$, varies linearly with the applied load, $P$. This is depicted in \Cref{fig:fad_basics_a} as the blue dashed line. Additionally, the load, $P_u$, at which ultimate plastic collapse occurs is shown as the brown dotted line.\\

For plastically deforming materials such as metals, a plastic zone develops around the crack tip. The elastic SIF is still a valid description of the crack tip state when the plastic zone is small. In this case, a $K$-dominated region in which the stress varies as $1/\sqrt{r}$ is still present around the crack tip. When the plastic zone becomes larger at higher loads, a $K$-dominated zone is no longer present. However, the state at the crack tip is still uniquely described by the $J$-integral, with a $J$-dominated zone being present around the crack tip. Although $K$ is strictly speaking not a valid crack tip parameter anymore, an effective SIF can be calculated by using the relation between $J$ and $K$ for linear elastic materials:
\begin{equation}
    K_{eff} = \sqrt{E'J},
\end{equation}
where $E'$ is the equivalent Young's modulus given as $E'=E/(1-\nu^2)$ for plane strain conditions. \Cref{fig:fad_basics_a} schematically depicts $K_{eff}$ as a function of the applied load. At small loads, where there still exists a $K$-dominated region, $K_{eff}$ is equal to $K$. At higher loads, plasticity starts to play a role, increasing $K_{eff}$ with respect to $K$. 

\subsection{Failure assessment}
\label{section:failure_assessment}

An FAD is an interaction diagram in which the driving forces for plastic collapse and LEFM are represented on the axes. This is depicted in \Cref{fig:fad_basics_b}. On the $x$-axis a load ratio, $L_r$, is represented that indicates the proximity to plastic collapse; $L_r=P/P_y$, where $P_y$ is the plastic yielding load. The $y$-coordinate represents the fracture ratio, given by
\begin{equation}
    K_r = \frac{K}{K_c},
    \label{eq:fractureratio}
\end{equation}
in which $K_c$ is the toughness of the material.\\ 

A failure assessment line (FAL) that governs the interaction between both principle failure criteria is plotted in the FAD of \Cref{fig:fad_basics_b}, denoted as $f(L_r)$. This curve represents the function $f=K_{eff}/K$ as a function of the applied load. Consequently, the FAD is a way to visualize the effective fracture driving force, $K_{eff}$. To assess whether a certain geometry with pre-existing crack fails under a certain load, an assessment point is plotted in the FAD by determining $L_r$ and $K_r$. Consider point $B$ in \Cref{fig:fad_basics_b}, which lies on the FAL $f(L_r)$. In this case, $K_r=K_I/K_c=f=K_I/K_{eff}$, which implies that $K_{eff}=K_c$, i.e.\ the effective stress intensity factor is equal to the material toughness and crack propagation occurs. Hence, a geometry and loading condition resulting in an assessment point lying on or outside of the FAL is unsafe. An assessment point lying within the FAL, such as point $A$ in \Cref{fig:fad_basics_b}, is considered safe. A safety factor for this assessment point can be calculated as $SF=OB/OA$, where $O$ is the (0,0) origin.\\

If the FAL is known, an FAD provides an easy-to-implement failure assessment method since only two parameters are involved; one based on plastic collapse and the other based on LEFM.

\subsection{Construction of the failure assessment line}

Current defect assessment codes provide three hierarchical options to determine the failure assessment line (FAL), which vary in complexity and accuracy \cite{bs7910}. In Option 3, the more general one, detailed elastic-plastic finite element analysis is conducted to calculate
\begin{equation}
    f_3(L_r) = \sqrt{\frac{J}{J_e}},
    \label{eq:fad3}
\end{equation}
over a range of $L_r$ values. Here, $J$ is the $J$-integral value obtained with an elastic-plastic analysis, while $J_e$ is the value of a purely elastic analysis. The obtained FAL depends on the material properties, geometry, and loading type, making it unsuitable for general application. Furthermore, the extensive finite element computations required make it a rarely adopted option in practical engineering applications. \\

In Option 2, approximate solutions of $J$ for materials obeying the Ramberg-Osgood relation are used \cite{Kumar1980}. By adopting a reference stress approach with an appropriate normalising load, the dependence of these solutions on the geometry and material properties is reduced \cite{Ainsworth1984,Ainsworth2003}. The normalizing load is given by
\begin{equation}
    L_r = \frac{P}{P_y},
    \label{eq:Lr}
\end{equation}
where $P_y$ is the plastic yield load of the flawed structure. The reference stress for a given load is then calculated as
\begin{equation}
    \sigma_\text{ref} = L_r \sigma_{y0},
    \label{eq:sigmaref}
\end{equation}
where $\sigma_{y0}$ is the initial yield stress of the material. The FAL  of Option 2, approximating \Cref{eq:fad3}, is given by
\begin{equation}
    f_2(L_r) = \left( \frac{E \varepsilon_\text{ref}}{\sigma_\text{ref}} + \frac{L_r^2 \sigma_\text{ref}}{2 E \varepsilon_\text{ref}} \right)^{-0.5},
    \label{eq:fad2}
\end{equation}
where $\varepsilon_\text{ref}$ is the reference strain corresponding to the reference stress, $\sigma_\text{ref}$. The construction of the Option 2 FAL only requires the stress-strain curve of the material, which defines the relationship between $\varepsilon_\text{ref}$ and $\sigma_\text{ref}$. Consequently, Option 2 is much easier to use in practice than Option 3. Since \Cref{eq:fad2} is derived based on the Ramberg-Osgood stress-strain relation, this equation is commonly fit to experimental data.\\

The actual sensitivity of \Cref{eq:fad2} to material behavior is limited. Therefore, defect assessment codes provide an Option 1, which is a conservative fit of \Cref{eq:fad2} based on data from various materials:
\begin{equation}
    f_1(L_r) = \left(1 + 0.5 L_r^2 \right)^{-0.5} \left( 0.3 + 0.7 \exp(-\mu L_r^6) \right),
\end{equation}
where
\begin{equation}
    \mu = \min \left( 0.001 \frac{E}{\sigma_{y0}}, 0.6 \right).
\end{equation}
The FAL of Option 1 is geometry and material independent, making it a popular approach.\\

The FALs are cut off at the point when plastic collapse structure would be expected, as shown in \Cref{fig:fad_basics_b}. The cut-off load is given by 
\begin{equation}
    L_{r,\max} = \frac{\sigma_{y0} + \sigma_{u}}{2 \sigma_{y0}},
    \label{eq:collapseload}
\end{equation}
where $\sigma_u$ is the ultimate strength of the material.

\subsubsection{Evaluation of $L_r$ and $K_r$}
\label{section:calculating_K_Py}

To calculate the ratios $L_r$ and $K_r$, it is necessary to determine the stress intensity factor $K$, as a function of the applied load, and the global yield load $P_y$. For simple geometries, solutions to $K$ and $P_y$ can be found in handbooks or codes. For more complex geometries, finite element analysis is required.\\

The stress intensity factor can be obtained through a linear elastic finite element calculation. The global yield load is the point at which yielding starts to affect the global load-displacement behavior of the structure. This value can be determined by conducting a limit load analysis using perfectly plastic material behavior. In such an analysis, the maximum load is reached when the uncracked ligament of the specimen is fully plastic. It is important to note that $P_y$ must take into account the location and size of the defects and cannot simply be calculated as the load at which the nominally applied stress reaches the yield stress.


\section{Elastic-plastic phase field fracture for hydrogen embrittlement}
\label{section:model}

An elastic-plastic phase field fracture model is employed for this analysis, coupled with a hydrogen transport formulation. The model is described in detail in Mandal et al.\ \cite{Mandal2024} and Wijnen et al.\ \cite{Wijnen2025}, and hence only key information is provided here, for the sake of brevity. Additional details are provided in \ref{section:appendix_model}, where the capabilities and behaviour of the model are explored further. Phase field fracture models build upon the thermodynamics of fracture, the energy balance first proposed by Griffith \cite{griffith1921vi} and later extended to elastic-plastic materials by Orowan \cite{orowan1949fracture}, and its predictions have been shown to agree with those of fracture mechanics theory. These aspects are not presented here, as they have been extensively discussed elsewhere (see, e.g., \cite{klinsmann2015assessment,Tanne2018,kristensen2021assessment} and Refs. therein). Nevertheless, \ref{section:flawsize} demonstrates, for the first time, that an elastic-plastic fracture model can naturally capture the role of elasticity, plasticity and crack length, revealing a good agreement with well-established fracture mechanics concepts (transition flaw size, FAD) and the ability to accurately predict fracture for any choice of crack length and material ductility.\\

The stress state in the material is governed by the linear momentum balance, given by
\begin{equation}
	\nabla \cdot \bm{\sigma}  = \mathbf{0}.
\end{equation}
A small strain formulation is adopted in which the strain is additively decomposed into an elastic and plastic part:
\begin{equation}
    \bm{\varepsilon} = \bm{\varepsilon}^e + \bm{\varepsilon}^p,
\end{equation}
The stress response of the material is given by
\begin{equation}
	\bm{\sigma} =  \kappa \left[ g(\phi) \left< \varepsilon^e_\text{vol} \right> - \left< -\varepsilon^e_\text{vol} \right>\right] \mathbf{I} + 2g(\phi)G \bm{\varepsilon}^e_\text{dev}
    \label{eq:stress}
\end{equation}
where $\varepsilon^e_\text{vol}=\text{trace}(\bm{\varepsilon}^e)$ is the volumetric elastic strain, $\bm{\varepsilon}^e_\text{dev}=\bm{\varepsilon}^e - \tfrac{1}{3}\varepsilon^{e}_\text{vol}\bm{I}$ is the deviatoric elastic strain tensor, $\kappa$ is the bulk modulus, $G$ is the shear modulus, $\left< \square \right>$ represent Macaulay brackets, and $g(\phi)$ is a degradation function. This degradation function is dependent on the order parameter $\phi$, which can be regarded as a damage indicator in the phase field fracture model. \Cref{eq:stress} is defined such that only the tensile hydrostatic and deviatoric parts of the stress are degraded, while the compressive hydrostatic part is not degraded \cite{Amor2009}.  \\

The evolution of the plastic strain is given by the normality flow rule:
\begin{equation}
	\dot{\bm{\varepsilon}}^p = \dot{\varepsilon}^p_\text{eq}  \frac{\partial f}{\partial \bm{\sigma}},
	\label{eq:plasticflow}
\end{equation}
where a yield surface with power-law hardening is adopted:
\begin{equation}
	f = \sigma_\text{eq} - g_p(\phi) \sigma_{y0}\left(1 + \frac{E \varepsilon^p_\text{eq}}{\sigma_{y0}}\right)^n,
	\label{eq:yieldsurface}
\end{equation}
subjected to the constraints
\begin{equation}
	\dot{\varepsilon}^p_\text{eq} \geq 0, \quad f \leq 0, \quad \dot{\varepsilon}^p_\text{eq}f=0.
	\label{eq:plasticconstraints}
\end{equation}
In \Cref{eq:yieldsurface}, $g_p(\phi)$ is used to degrade the plastic yield surface as a function of the order parameter. Furthermore, $\varepsilon^p_\text{eq}=\sqrt{\tfrac{2}{3} \bm{\varepsilon}^p_\text{dev} : \bm{\varepsilon}^p_\text{dev}}$ denotes the equivalent plastic strain.\\

A crack is represented by an order parameter $\phi$. A value of $\phi=0$ represents intact material, while a value of $\phi=1$ represents a fully damaged material. The evolution of the order parameter is governed by the following partial differential equation
\begin{equation}
	\frac{G_c^\text{H}(C)}{\ell} \left( \phi - \ell^2 \nabla^2 \phi \right) = -\frac{dg}{d\phi} (\mathcal{H} + \beta \psi^{p} ),
	\label{eq:localbalance_phi}
\end{equation}
in which $G_c(C)^\text{H}$ is a (hydrogen-dependent) critical energy release rate, also known as material toughness, and $\ell$ is a length scale which regularizes the order parameter. The right side of \Cref{eq:localbalance_phi} represents the crack driving force. Here, $\mathcal{H}$ is the irreversible elastic crack driving force, given by
\begin{equation}
	\mathcal{H} = \max_{\tau\in[0,t]}\left( \frac{\kappa}{2} \left< \varepsilon^e_\text{vol} \right>^2 + G \bm{\varepsilon}^e_\text{dev} : \bm{\varepsilon}^e_\text{dev} \right).
\end{equation}
In addition, the plastic dissipated energy can be calculated through
\begin{equation}
	\psi^p = \frac{\sigma_{y0}^2}{E(n+1)} \left( 1 + \frac{E \varepsilon^p_\text{eq}}{\sigma_{y0}} \right)^{n+1} - \frac{\sigma_{y0}^2}{E(n+1)}.
	\label{eq:plasticenergy}
\end{equation}
A parameter $\beta$ is adopted in \Cref{eq:localbalance_phi} such that only a part of the plastic dissipation contributes to damage. This parameter is taken equal to $\beta=0.1$ as experiments show that 90\% of the plastic work is dissipated into heat \cite{Taylor1934}, and thus not available to drive crack growth.\\

Finally, the elastic and plastic degradation functions are respectively given by
\begin{equation}
    g(\phi) = (1 - \phi)^2
\end{equation}
and 
\begin{equation}
    g_p(\phi) = \beta g(\phi) - \beta + 1,
\end{equation}
where the latter is adopted to ensure thermodynamic consistency with \Cref{eq:localbalance_phi} \cite{Mandal2024}.

\subsection{Hydrogen diffusion}

The diffusion of atomic hydrogen through the material is described by
\begin{equation}
	\frac{\partial C}{\partial t} = \nabla \cdot \left( D \nabla C - \frac{DC}{RT} V_h \nabla \sigma_h \right),
\end{equation}
where $C$ is the diffusible hydrogen concentration, $D$ is the apparent diffusivity coefficient, $V_h$ is the partial molar volume of hydrogen, and $\sigma_h= \text{trace}(\bm{\sigma})/3$ is the hydrostatic stress. The local hydrogen concentration degrades the local critical energy release rate through a phenomenological degradation law \cite{Mandal2024}:
\begin{equation}
	G_c^\text{H}(C) = f(C) G_{c} = 
        \left[ f_\text{min} 
	+ \left(1-f_\text{min}\right) 
	\exp(-q_1 C^{q_2})
	\right] G_{c},
 \label{eq:degradationfunction}
\end{equation}
where $G_{c}$ is the critical energy release rate of the material in air and $f_\text{min}$, $q_1$, and $q_2$ are other material parameters that are calibrated against toughness versus hydrogen content data, as elaborated below.

\section{Virtual FAD of a SENT specimen}
\label{section:sent}

In this section, a single-edge notched tension (SENT) specimen with homogeneous properties is considered. This relatively simple geometry allows us to demonstrate how the adopted phase field fracture method compares with traditional FAD assessment, validating capabilities of the phase field fracture method to accurately predict elastic-plastic fracture (Section \ref{section:sent_fad}). Furthermore, the SENT specimen will be used to study the effect of hydrogen on the shape of the FAD (Section \ref{section:sent_fad_H}).

\subsection{FAD and model comparison}
\label{section:sent_fad}

\begin{figure}
    \begin{subfigure}{\linewidth}
        \centering 
        \includegraphics[width=.8\linewidth]{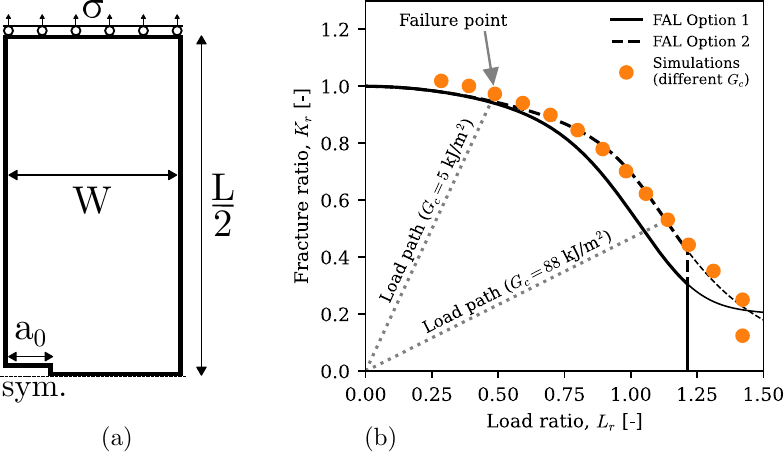}
        \phantomcaption \label{fig:sent_sketch}
        \phantomcaption \label{fig:sent_fad}
    \end{subfigure}
    \caption{Failure assessment diagram for a SENT specimen. (a) Sketch of the SENT geometry. (b) Comparison between a virtual failure assessment line (FAL), obtained by model predictions with different fracture toughness values, with traditional FALs according to standards.}
    \label{fig:sent}
\end{figure}

The SENT specimen considered is depicted in \Cref{fig:sent} and has a width $W$ of 5 mm, a length (height) $L$ of 50 mm, and an initial crack length $a_0$ of 1.5 mm. These dimensions comply with the standards for fracture toughness testing \cite{bs8571}. The $a/W$ ratio is chosen such that failure is driven by crack propagation and a $J$-dominant field is present around the crack tip. For large ratios $a/W$, large plastic deformations occur in the entire remaining ligament, resulting in the absence of a $K$ or $J$ field. On the other hand, for very small values of $a/W$ failure is governed by the material strength. A discussion on crack lengths and corresponding transition in failure modes is presented in \ref{section:flawsize}, in which a connection between FADs and transition flaw analysis is established.\\

Solutions for the stress intensity factor and limit load are available for a SENT specimen \cite{Anderson2017,Milller1988}. The stress intensity factor is calculated by equation \Cref{eq:sif}, with the geometry-dependent function given as \cite{Anderson2017}
\begin{equation}
    f\left(\frac{a}{W}\right) = \frac{\sqrt{2 \tan [\pi a / (2W) ]}}{\cos[\pi a/(2W)]} \left[ 0.752 + 2.02\left( \frac{a}{W} \right) + 0.37\left( 1 - \sin \frac{\pi a}{2W} \right)^3 \right].
    \label{eq:faw_sent}
\end{equation}
The yield load of the specimen, used to calculate the normalized load through \Cref{eq:Lr}, is given by \cite{Milller1988} 
\begin{equation}
    P_y = 1.155 \, B \, W \, \sigma_{y0} \left(1 - \frac{a}{W} \right).
    \label{eq:Py_sent}
\end{equation}
The validity of \Cref{eq:faw_sent,eq:Py_sent} are verified through finite element calculations, as described in \Cref{section:calculating_K_Py}.\\

The fracture toughness $K_c$ (required to calculate the fracture ratio, see \Cref{eq:fractureratio}) is related to the critical energy release rate $G_{c}$ (key input to the phase field fracture model) through
\begin{equation}
    K_c = \sqrt{E' G_c}.
\end{equation}
To span multiple regions of the failure assessment diagram, simulations with varying $G_{c}$ were conducted. The elastic and plastic properties were kept constant and are presented in \Cref{table:parameters_sent}. The value of $\ell$ in each simulation is chosen such that the ductility ratio remains constant at $r_y = 1.5$ (see \ref{section:appendix_model}). Each simulation represents a different loading path in the FAD. These loading paths are straight lines starting at the origin of the FAD. In \Cref{fig:sent_fad}, the loading paths associated with $G_{c}=5$ kJ/m$^2$ and $G_{c}=88$ kJ/m$^2$ are shown. As the applied load increases, the simulations progress along these paths. The load at which the crack propagates is marked with an orange circle. For each simulation, this failure point is plotted in \Cref{fig:sent_fad}. The contour that is obtained by the simulation failure points can be regarded as a \emph{virtual} failure assessment line.\\ 

\begin{table}[!tb]
\label{table:parameters_sent}
\caption{Model parameters adopted for the SENT specimen.}
\centering
\begin{tabular}{l|l|l|l|l|l|l|l}
$E$ {[}GPa{]} & $\nu$ {[}-{]} & $\sigma_{y0}$ {[}MPa{]} & $n$ {[}-{]} & $D$ {[}mm$^2$/s{]} & $f_{min}$ {[}-{]} & $q_1$ {[}-{]} & $q_2$ {[}-{]} \\ \hline
200                            & 0.3                            & 800                                   & 0.1                             & $4.5 \cdot 10^{-4}$                      & 0.65              & 30            & 1            
\end{tabular}
\end{table}

In addition, FAL options 1 and 2 are displayed in \Cref{fig:sent_fad}, where a Ramberg-Osgood relation between $\varepsilon_\text{ref}$ and $\sigma_\text{ref}$ is assumed for the construction of the Option 2 FAL. It can be seen that the failure points of the simulations correspond nearly exactly with the Option 2 FAL, meaning that the failure load predicted by both methods is in excellent agreement. Simulation failure points that lie beyond the standard FALs would mean that the model predicts failure later than with the FAD methods. On the other hand, simulation failure points that lie inside the standard FALs would mean that the model predicts failure earlier than the FAD methods. The excellent agreement between the virtual FAL and the Option 2 FAL, for the simple case of a homogeneous SENT, demonstrates that the adopted elastic-plastic phase field fracture method is in good agreement with the traditional fracture mechanics theory upon which the FAD assessment is based.\\

One can also see that the plastic collapse cut-off load of the standard FALs is lower than that of the virtual FAL. The standard plastic collapse load, given by \Cref{eq:collapseload}, is based on the flow stress of the material, $\sigma_f = 0.5(\sigma_{y0}+\sigma_u$). However, since a small strain approximation is adopted in the material, the load in a simulation where no crack growth occurs keeps increasing and saturates towards a load ratio of $\sigma_u/\sigma_{y0}$. It is expected that finite deformation effects would reduce the plastic collapse load obtained with the simulations, but play a negligible role in the other regimes of the FAD \cite{tvergaard1992relation}. 

\subsection{Hydrogen-affected FAD}
\label{section:sent_fad_H}

To study the effect of hydrogen on the shape of the FALs, the simulations are repeated for a SENT specimen exposed to hydrogen. On both the left and right faces of the geometry in \Cref{fig:sent_sketch} a hydrogen concentration of $C=0.17$ wppm is considered. According to Sievert's law, this level of hydrogen corresponds to a hydrogen pressure of approximately 5 MPa. The adopted hydrogen degradation parameters (\Cref{eq:degradationfunction}) are given in \Cref{table:parameters_sent}. A slow loading rate was adopted such that steady state conditions were approached when the crack was not propagating.\\

\begin{figure}[!tb]
    \begin{subfigure}{\linewidth}
        \includegraphics[width=\linewidth]{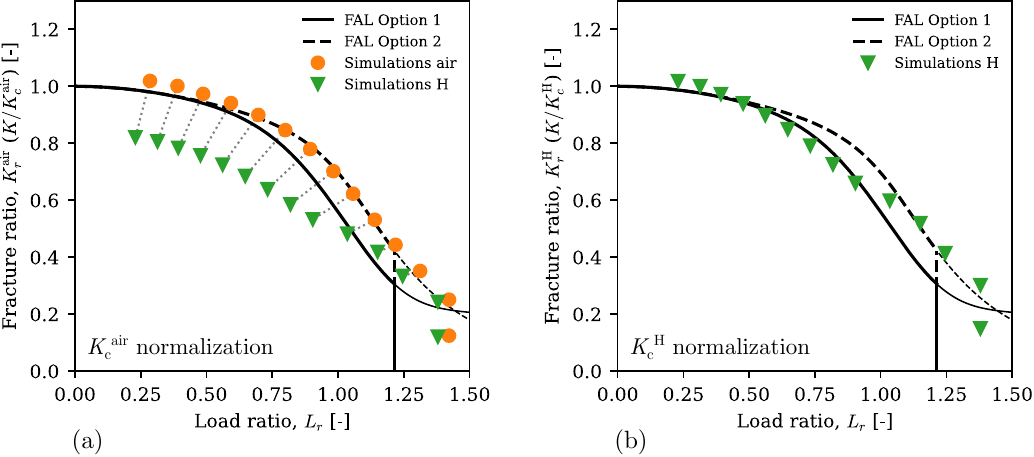}
        \phantomcaption \label{fig:fad_H_a}
        \phantomcaption \label{fig:fad_H_b}
    \end{subfigure}
    \caption{FADs for a SENT specimen exposed to hydrogen. (a) FAD constructed using the fracture toughness of the material in air to determine the fracture ratio. (b) FAD constructed using the fracture toughness of the material in hydrogen to determine the fracture ratio.}
    \label{fig:fad_H}
\end{figure}

The results of this analysis are presented in \Cref{fig:fad_H}. In \Cref{fig:fad_H_a}, the failure points in hydrogen are plotted on the same FAD as the failure points in air. This means that the fracture ratio, $K_r^\text{air}$, for all data points is calculated by normalising with respect to the critical stress intensity factor of the material in air, $K_c^\text{air}$. The results show that hydrogen reduces the failure load of the specimen. This effect is most pronounced in the brittle regime, corresponding to the upper-left region of the FAD. For points failing in the ductile regime, i.e.\ those with a load ratio close to or above 1, the reduction in load is less pronounced. In cases where failure occurs due to plastic collapse, the effect of hydrogen is nearly negligible.\\

To evaluate the safety of a flawed structure exposed to hydrogen, the fracture properties of the material in a hydrogen environment need to be known. In this case, a simple adaption to the FAD method can be made by calculating the fracture ratio, $K_r^\text{H}$, by normalizing with respect to the fracture toughness in hydrogen, $K_c^\text{H}$ \cite{Martinez2019,Pluvinage2021}. For the SENT specimen considered here, this fracture toughness is based on the degraded critical energy release rate (\Cref{eq:degradationfunction}) at the hydrogen concentration that is applied to the faces of the specimen ($C=0.17$ wppm). \Cref{fig:fad_H_b} presents the resulting FAD, incorporating hydrogen degradation effects. A reasonable agreement between the virtual FAL and standard FALs is observed. At both extremes of the FAD, representing purely brittle or ductile failure, the virtual FAL aligns closely with the Option 1 FAL. However, in the transition regime between brittle to ductile failure ($0.6<Lr<0.9$), the virtual FAL shows a small dip, resulting in a slightly different shape compared to the standard FALs and the virtual FAL in air. Within this regime, the simulation failure points roughly coincide with the more conservative Option 2 FAL. This indicates that adopting Option 2 would still adequately identify safe operating loads in this scenario.

\section{FADs of pipeline welds}
\label{section:pipelines}

In this section, multiple defects around a pipeline weld are simulated, taking into account heterogeneous material properties that are obtained from a welding simulation. FADs are constructed based on global material behavior. Therefore, by accounting for local material properties, the model serves as a reference for realistic pipeline welds. This approach allows for the assessment of assumptions used in defect assessment with FADs by comparing them with simulation results.

\subsection{Welding model results and material data}

\begin{figure}[!tb]
    \begin{subfigure}{\linewidth}
        \centering
        \includegraphics[width=\linewidth]{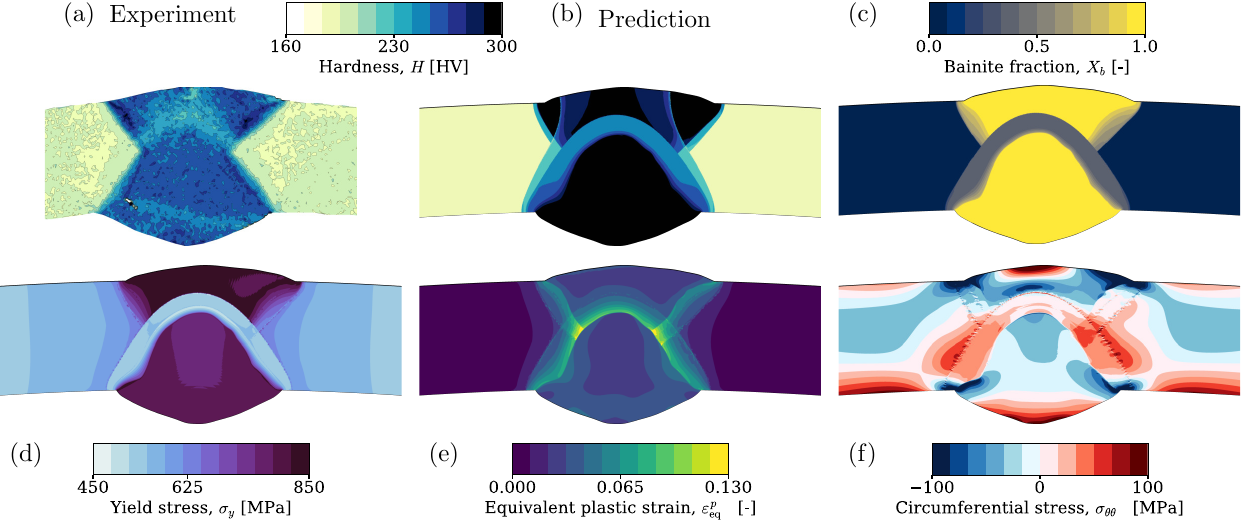}
        \phantomcaption \label{fig:weldresults_hardness_exp}
        \phantomcaption \label{fig:weldresults_hardness}
        \phantomcaption \label{fig:weldresults_xb}
        \phantomcaption \label{fig:weldresults_yield}
        \phantomcaption \label{fig:weldresults_strain}
        \phantomcaption \label{fig:weldresults_stress}
    \end{subfigure}
    \caption{Welding model results. Hardness maps (a) obtained experimentally and (b) predicted by the model. (c) Bainite fraction predicted by the simulation. (c) Yield stress. (c) Equivalent plastic strain. (f) Residual circumferential stress.}
    \label{fig:weldresults}
\end{figure}

An X60 pipeline is considered in the welding model. 
The geometry of the weld was reconstructed based on available macrographs. The pipeline had an outer diameter of 762 mm, with a wall thickness of 12.7 mm.  A welding simulation was conducted, which predicted the phase and property distributions and the residual stresses within the weld. For details of this welding simulation, the reader is referred to Wijnen et al.~\cite{Wijnen2025}. \Cref{fig:weldresults} presents a selection of results obtained with the welding simulation, including yield stress, equivalent plastic strain, and residual circumferential stress. \Cref{fig:weldresults_hardness_exp,fig:weldresults_hardness} show, respectively, the experimental and predicted hardness maps, which are in good agreement. The weld metal and narrow zones adjacent to the weld metal, which are termed heat affected zones (HAZ), exhibit a significantly higher hardness than the base metal. Such hard regions are typically more susceptible to failure. In \Cref{fig:weldresults_xb}, the bainite phase fraction of the underlying microstructure is shown. The microstructure in the weld metal and the HAZ consists of a percentage of bainite ranging from 30\% to 100\%. In contrast, the base metal is composed of a mixture of ferrite and pearlite. The bainite present in the weld metal and HAZ microstructures is the primary reason for the high hardness in these regions.\\

\begin{figure}[h]
    \begin{subfigure}{\linewidth}
        \centering
        \includegraphics[width=\linewidth]{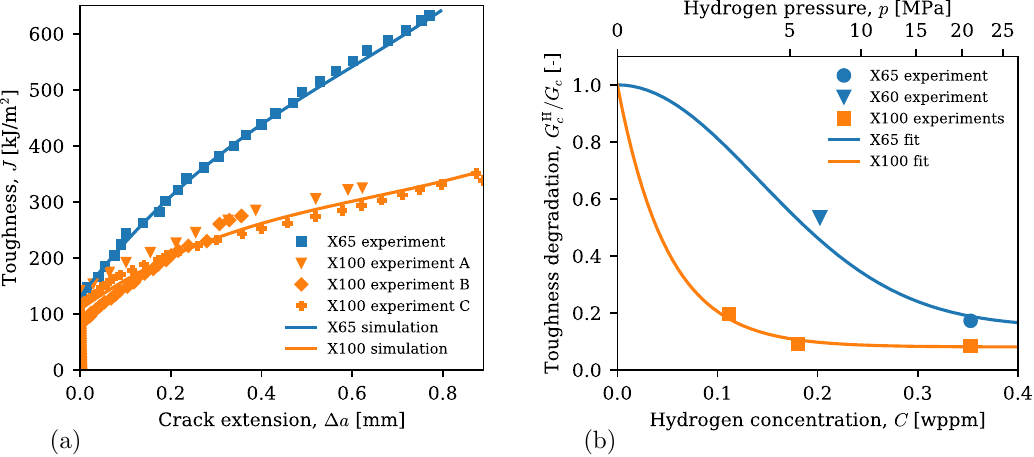}
        \phantomcaption \label{fig:properties_Jcurves}
        \phantomcaption \label{fig:properties_degradation}
    \end{subfigure}
    \caption{Fracture toughness data. (a) Experimental $J$-curves and model fits obtained by boundary layer simulations. (b) Fit of the fracture toughness degradation laws through experimental data.}
    \label{fig:properties}
\end{figure}

The parameters of the fracture model are based on experimental crack growth resistance curves ($J$-curves), obtained by Ronevich and coworkers \cite{Ronevich2016b,Ronevich2021}. The behavior of ferrite and pearlite is based on data of the base metal of an X65 steel, which has a ferritic-pearlitic microstructure and is expected to exhibit similar mechanical properties as an X60 grade steel. The fracture behavior of the bainite phase is based on data of an X100 pipeline steel, which has a bainitic microstructure. The $J$-curves of both materials are presented in \Cref{fig:properties_Jcurves}, together with the predicted $J$-curves obtained through boundary layer simulations. The $J$-curves of both materials start at a similar toughness. However, the X65 steel has a significantly higher slope than the X100 steel. The model is able to accurately capture the experimental $J$-curves.

\Cref{fig:properties_degradation} depicts the adopted toughness degradation laws (\Cref{eq:degradationfunction}). For the X100 grade, a sufficient number of data points is available to obtain a unique fit. In contrast, only a single data point is available for the X65 steel. 
Therefore, this data is complemented with an X60 steel with a similar microstructure \cite{Hoover1981}.
It can be seen that bainite undergoes more severe fracture toughness degradation when exposed to hydrogen.

\subsection{FAD assessment of a pipeline in air}
\label{section:pipeline_air}

\begin{figure}[h]
    \centering
    \includegraphics[width=.7\linewidth]{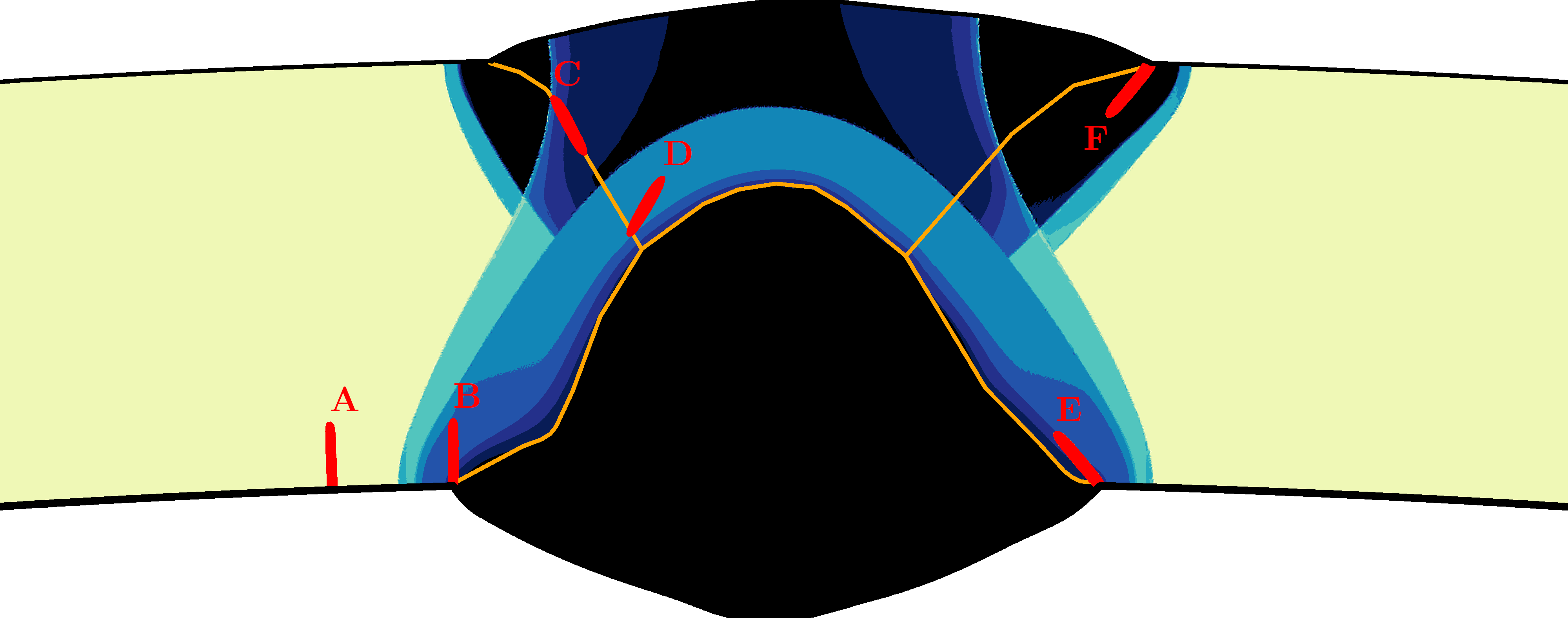}
    \caption{Locations of all the defects considered in this study. Defects with a length of 2mm are depicted.}
    \label{fig:initialdefects}
\end{figure}

Initial defects in various locations were considered, which are depicted in \Cref{fig:initialdefects}, and are denoted by the characters \emph{A} to \emph{F}. These represent typical locations where defects in pipelines are commonly observed. For each defect location, two defect lengths were examined, namely, 2 mm and 4 mm. Each simulation took only a single defect into account, with no simulations involving multiple defects.\\

First, we consider simulations that were conducted assuming homogeneous properties throughout the entire pipeline, corresponding to those of the base metal. The heterogeneity introduced during the welding process was not considered. For each crack, we use finite element calculations to determine the SIF, $K$, as a function of the pressure, and the global yield load, $P_y$, as detailed in \Cref{section:calculating_K_Py}. Cracks \emph{A} and \emph{B} experience Mode I loading, while the loading of cracks \emph{C} through \emph{F} has a mixed-mode character. For mixed-mode conditions, $K$ is calculated as the SIF associated with the maximum energy release rate criterion. The pressure in the pipeline was slowly increased by applying a radial displacement to its inner surface. The actual pressure in the pipeline was calculated using the hoop stress and the equation for thin-walled cylinders: $p=2\sigma_{\theta\theta}t/(2r)$, where $t$ denotes thickness and $r$ radius.\\ 

\begin{figure}[!tb]
    \begin{subfigure}{\linewidth}
        \includegraphics[width=\linewidth]{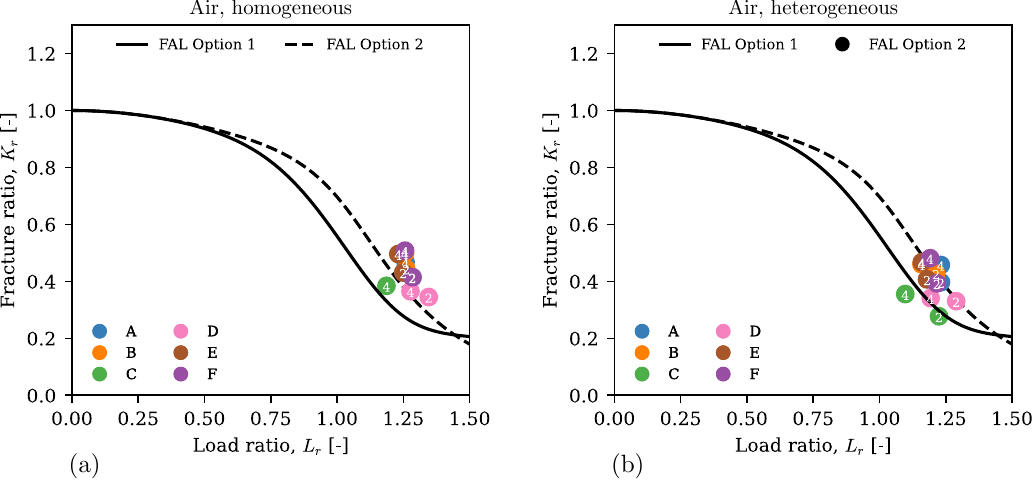}
        \phantomcaption \label{fig:pipe_fad_homog}
        \phantomcaption \label{fig:pipe_fad_heter}
    \end{subfigure}
    \caption{FADs of the pipeline weld in air. (a) Predicted failure points for a weld with homogeneous properties equal to that of the base metal. (b) Failure points for a heterogeneous weld.}
    \label{fig:pipe_fad}
\end{figure}

The failure loads for the different crack locations are plotted on the FAD in \Cref{fig:pipe_fad_homog}. The initial crack length is indicated within each marker in millimeters. All failure points lie relatively close to the standard Option 2 FAL, with several points below and above the line. The small scatter can be attributed to the more complex geometry compared to the SENT specimen considered in \Cref{section:sent}. Nevertheless, the Option 2 FAL gives a reasonable prediction of the failure load. In addition, all points lie above the more conservative Option 1 FAL. This indicates that the FAL adequately predicts safe operating conditions in this situation.\\

Next, simulations were performed by incorporating heterogeneous properties in the weld region, as obtained from the welding model (\Cref{fig:weldresults}). In practice, however, the exact local material properties in the weld and heat affected zones are generally unknown. Testing specific regions in the weld and heat affected zones is challenging and thus, frequently only base metal properties are known. Therefore, we construct the FAD using driving forces normalized based on the base metal properties. The fracture ratio, $K_r$, is calculated using the fracture toughness, $K_c$, of the base metal. The evaluation of the load ratio, $L_r$, is conducted based on the value of $P_y$ obtained using the homogeneous properties of the base material. The failure data points for the simulations incorporating heterogeneity are shown in \Cref{fig:pipe_fad_heter}. Since the same $K_c$ and $P_y$ as for the homogeneous case are used to normalize the FAD axes, the failure points lie on the same loading path. For most cracks, failure occurred slightly earlier in the heterogeneous case compared to the homogeneous case, so their failure point is located lower on the loading path. However,  the differences are small. Although the yield strength of the material is higher in the weld, the initial fracture toughness in air is similar throughout the entire pipeline, resulting in only minor differences. The failure points of the 2 mm and 4 mm defects at location \emph{C} and the 4 mm defect at location \emph{D} lie approximately on the Option 1 FAL. The above results suggest that the assessment of defects using an FAD constructed using base metal properties is justified for this particular weld.

\subsection{FAD assessment of a pipeline in hydrogen}

The same simulations were performed but with the pipeline subjected to hydrogen pressure. In this situation, the hydrogen diffuses through the metal, leading to a reduction in fracture toughness. A hydrogen concentration was applied to the inner surface of the pipeline, corresponding to the applied pressure through Sievert's law, given by $C^*_\text{inner}=S\sqrt{p}$, where $S$ is the hydrogen solubility of the material, taken as 0.077 wppm$\cdot$MPa$^{-0.5}$, as measured for pipeline steels. A zero hydrogen concentration was prescribed to the outer surface of the pipeline, resulting in a gradient in the hydrogen concentration. Furthermore, a very slow loading rate was adopted so that a steady-state condition was present when no crack propagation occurred.\\ 

\begin{figure}[h]
    \begin{subfigure}{\linewidth}
        \centering
        \includegraphics[width=\linewidth]{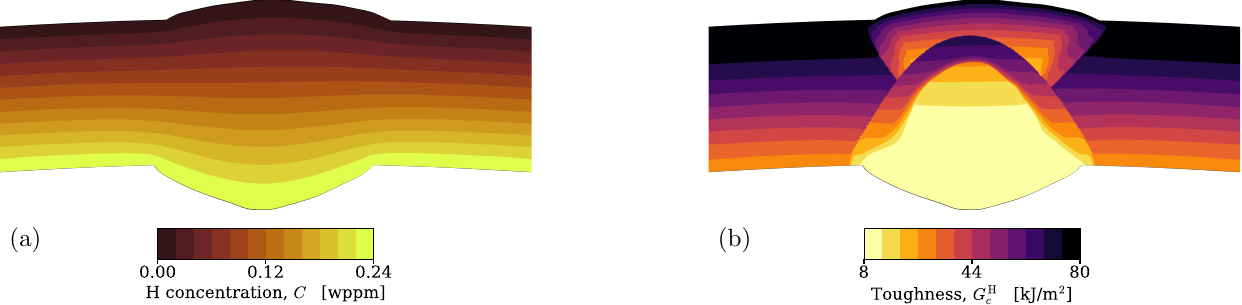}
        \phantomcaption \label{fig:distribtuion_hydrogen}
        \phantomcaption \label{fig:distribution_toughness}
    \end{subfigure}
    \caption{The effect of hydrogen on fracture toughness in the weld. (a) Hydrogen distribution at a pressure of 10 MPa. (b) Fracture toughness distribution for the hydrogen distribution of (a).}
    \label{fig:distribution}
\end{figure}

\Cref{fig:distribtuion_hydrogen} displays the hydrogen concentration at a pressure of approximately 10 MPa. The corresponding toughness distribution is shown in \Cref{fig:distribution_toughness}. The toughness in the weld metal and HAZ regions is substantially more degraded compared to the base metal region. Moreover, since the hydrogen concentration is higher close to the inner surface of the pipe, the toughness in this region of the weld is lower than in the weld region close to the outer surface.\\

\begin{figure}[!tb]
    \begin{subfigure}{\linewidth}
        \includegraphics[width=\linewidth]{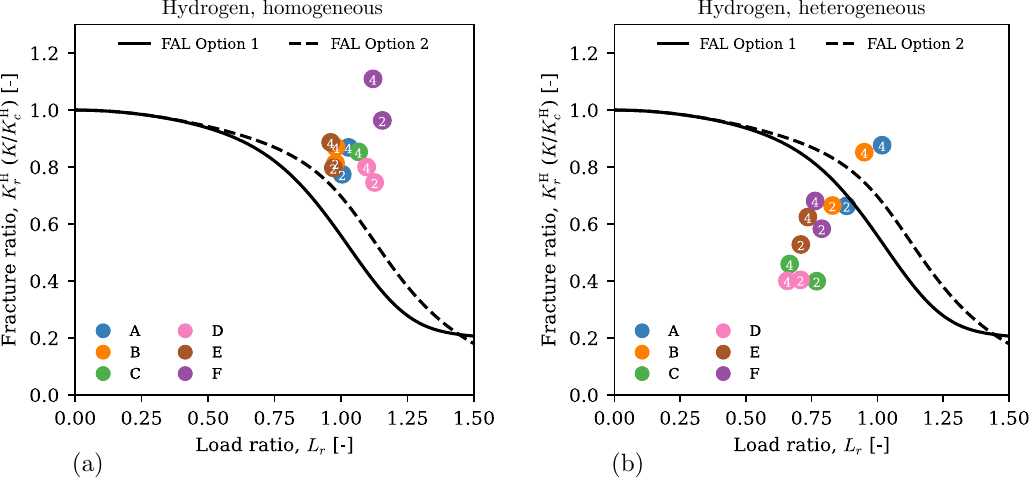}
        \phantomcaption \label{fig:pipe_fad_H_homog}
        \phantomcaption \label{fig:pipe_fad_H_heter}
    \end{subfigure}
    \caption{FADs of the pipeline weld subjected to hydrogen. The fracture ratio is determined by normalizing with the degraded fracture toughness of the base metal. (a) Predicted failure points for a weld with homogeneous properties equal to that of the base metal. (b) Failure points for a heterogeneous weld.}
    \label{fig:pipe_fad_H}
\end{figure}

First, the homogeneous pipeline, consisting entirely of base metal, is considered. In the same way as for the SENT specimen in \Cref{section:sent_fad_H}, the FAD is adapted for hydrogen (i.e., the fracture ratio, $K_r^\text{H}$, is normalized using the degraded fracture toughness). For this reason, it is assumed that the fracture toughness is known for relevant hydrogen concentrations. For the considered base metal properties, the toughness degradation is represented by the curve labeled X65 in \Cref{fig:properties_degradation}. The degraded $K_c^\text{H}$ is determined using the hydrogen concentration applied to the inner surface of the pipeline. Note that this can be considered conservative, as the hydrogen concentration decreases towards the outer surface of the pipeline. The failure points of all the considered cracks for the pipeline subjected to hydrogen pressure are shown in \Cref{fig:pipe_fad_H_homog}. Since the hydrogen concentration at the inner surface is dependent on the pressure, $K_c^\text{H}$ also changes with increasing pressure. As a result, the load path in the FAD is no longer a straight line. 
All the failure points lie above the Option 2 FAL, indicating that no simulation failed at an operating pressure considered safe according to an FAD assessment.\\

Next, we consider the pipeline exhibiting heterogeneous properties. Similar to \Cref{section:pipeline_air}, we first construct the FAD assuming only the base metal properties are known. This means that the fracture ratios are determined using $K_c^\text{H}$, which is degraded based on the degradation law of the base metal. The results are depicted in \Cref{fig:pipe_fad_H_heter}. Many of the simulations failed earlier than predicted by the standard FALs. Especially, defects at locations \emph{C}, \emph{D}, \emph{E}, and \emph{F} failed at pressures that are still considered safe by the FALs. This can be attributed to the much more severe toughness degradation in the bainitic weld metal, where these defects are located, compared to the ferritic-pearlitic base metal, as can be seen in \Cref{fig:properties_degradation}. These results reveal the risks associated with assessing defects in welds using FADs constructed based on base metal properties.\\

\begin{figure}[!tb]
    \begin{subfigure}{\linewidth}
        \includegraphics[width=\linewidth]{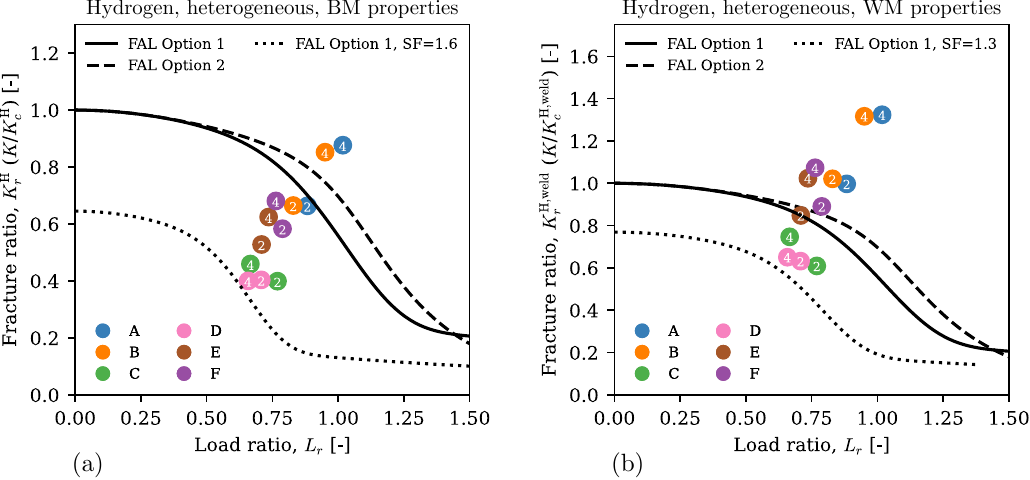}
        \phantomcaption \label{fig:pipe_fad_H_heter_sf}
        \phantomcaption \label{fig:pipe_fad_H_heter_corrected}
    \end{subfigure}
    \caption{FADs with safety factors of the pipeline subjected to hydrogen. (a) The fracture ratio is determined by normalizing with the degraded fracture toughness of the base metal. (b) The fracture ratio is determined by normalizing with the degraded fracture toughness of the weld metal.}
\end{figure}

A safety factor can be calculated, as explained in \Cref{section:failure_assessment}, for which all failure points lie above the FAL. \Cref{fig:pipe_fad_H_heter_sf} shows the FAD with a safety factor of 1.6 applied. Consequently, for this specific weld, a safety factor of a minimum safety factor of 1.6 is necessary to identify safe operating pressure limits. \\

The above results suggest that the specific properties of the weld and HAZs need to be known to safely apply FADs for defects in hydrogen environments. The narrowness of these regions, specifically the HAZ, makes obtaining these properties challenging. Nevertheless, several experimental studies to quantify the hydrogen-dependent fracture toughness have been conducted \cite{Chatzidouros2011,Agnani2023,Ronevich2021,Bortot2024,Martin2022,Chowdhury2024b}.\\

According to the BS7910 standard \cite{bs7910}, a conservative FAD can be constructed by calculating the load ratio assuming tensile properties that are the lowest of the base metal, weld metal, or HAZ. In addition, to have a conservative estimate of the fracture ratio, the fracture toughness needs to represent the lowest toughness microstructure present in the component. As observed in \Cref{fig:weldresults_yield}, the lowest yield stress corresponds to the base metal. The lowest toughness regions, in the presence of hydrogen, in the weld exhibit a fully bainitic microstructure, as can be seen in \Cref{fig:weldresults_xb,fig:distribution_toughness}. This means that the conservative FAD needs to be constructed using the degraded $K_c^\text{Hweld}$ of a fully bainitic material (the curve labeled X100 in \Cref{fig:properties_degradation}), while the yield strength of the base metal to determine $P_y$. \Cref{fig:pipe_fad_H_heter_corrected} displays the resulting FAD. Compared to the FAD constructed using base metal properties (\Cref{fig:pipe_fad_H_heter}), the failure points of the defects in the weld (\emph{C}, \emph{D}, \emph{E}) are located closer to the FALs. However, those of locations \emph{C} and \emph{D} still lie below both FALs. This indicates that the constructed FAD is still non-conservative for the considered weld, exhibiting heterogeneous properties and exposed to hydrogen. To ensure a safe assessment of the weld using the FAD, a safety factor needs to be taken into account. The dotted line in \Cref{fig:pipe_fad_H_heter_corrected} displays the Option 1 FAL, using a safety factor of $SF=1.3$. In this case, all failure points are located above the FAL. A safety factor of at least 1.3 is thus required for a safe assessment of the considered weld.\\

These findings highlight the importance of characterising the material properties in different regions of a weld. Ideally, FADs should be constructed using material properties specific to the region where the defect is located. To evaluate all defects, a conservative FAD can be constructed, requiring the properties of the most susceptible region. However, even in this scenario, applying a suitable safety factor is required. 

\subsection{Failure pressures}

\begin{figure}[!tb]
    \centering
    \includegraphics[width=.5\linewidth]{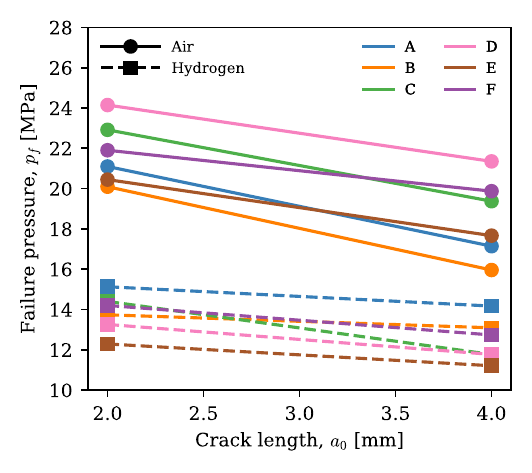}
    \caption{Failure pressures for defects at the considered locations in the weld.}
    \label{fig:pipe_pressures}
\end{figure}

While FADs have been used to analyze flawed pipelines, they do not provide information on actual failure pressures. Therefore, the failure pressures of the defects in both air and hydrogen environments, obtained from simulations incorporating heterogeneous properties, are presented in \Cref{fig:pipe_pressures}. The results reveal a distinct trend in air, where the failure pressures decrease with increasing initial crack size. In contrast, this trend is less pronounced in hydrogen, indicating a reduced sensitivity to crack size. These findings suggest that in hydrogen environments, small defects can be nearly as critical as larger defects, highlighting that even minor defects can significantly compromise the structural integrity of pipelines.\\


Notable reductions in failure pressures are observed when hydrogen is used instead of air. For the 2 mm defects, the average decrease in failure pressures is 7.9 MPa. The most pronounced reduction of 10.9 MPa occurs for the 2 mm defect at location \emph{D}, whereas the least pronounced reduction of 6.0 MPa is observed for the defect at location \emph{A}. This indicates that cracks within the weld region are significantly more susceptible to hydrogen embrittlement compared to cracks in the base metal. For the 4 mm defects, the average decrease in failure pressures is 6.1 MPa.

\section{Conclusions}

We present the first generation of \emph{Virtual} failure assessment diagrams (FADs), to address the challenges in fitness-for-service assessments of hydrogen transport pipelines. In particular, we combine state-of-the-art weld process modelling and phase field-based elastic-plastic deformation-diffusion-fracture simulations to predict weld integrity. The model delivered an excellent agreement with standardised failure assessment lines (FAL) for the case of an homogeneous material, with and without hydrogen exposure. This validated model was then used to incorporate the role of the two key uncertainties driving weld integrity: microstructural heterogeneity (hard, brittle HAZs) and residual stresses. The main findings are:
\begin{itemize}
    \item Under the influence of hydrogen, an adequate assessment \emph{of base metal behaviour} can still be made using an FAD when the fracture toughness used for constructing the FAD is taken at the specific hydrogen concentration.
    \item For welds exhibiting homogeneous properties, an FAD accurately identifies safe operating pressures. This holds even under the influence of hydrogen, given that the fracture toughness of the material at relevant hydrogen concentrations is known and used in the FAD construction.
    \item For a heterogeneous weld that is not exposed to hydrogen, an adequate defect assessment can be performed using an FAD, even when the exact weld properties are not known.
    \item When a heterogeneous weld is exposed to hydrogen, a standard FAD assessment is non-conservative, even when the properties of the most susceptible region in the weld are used to construct the FAD. For the specific weld considered in this study, a safety factor of at least 1.3 was required to identify safe operating pressures using the FAD assessment when the FAD was constructed based on weld metal properties. If base metal properties are used to construct the FAD, a minimum safety factor of 1.6 is necessary.
\end{itemize}

The present work enables quantifying the uncertainties driving the failure of welds in hydrogen transmission and storage components, linking state-of-the-art mechanistic models with simpler fitness-for-service tools, delivering appropriate safety factors and providing a pathway to safely design components for the hydrogen economy. Moreover, the framework is universal and also allows assessing the structural integrity of welded pipelines beyond hydrogen transmission applications.

\section*{Acknowledgments}
\label{Acknowledge of funding}

\noindent The authors acknowledge financial support from EPRI through the R\&D project ``Virtual Testing of hydrogen-sensitive components''. E.\ Mart\'{\i}nez-Pa\~neda acknowledges financial support from UKRI's Future Leaders Fellowship programme [grant MR/V024124/1] and from the UKRI Horizon Europe Guarantee programme (ERC Starting Grant \textit{ResistHfracture}, EP/Y037219/1).

\section*{Data availability statement}

The ABAQUS subroutines used in this work will are available at \url{https://mechmat.web.ox.ac.uk/codes}. 

\appendix
\section{Homogeneous model behavior}
\label{section:appendix_model}

\begin{figure}[!tb]
    \centering 
    \includegraphics[width=.5\linewidth]{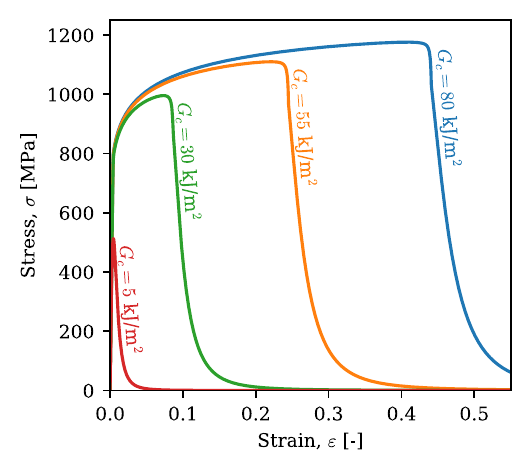}
    \caption{Stress-strain curves predicted by the model for different $G_c$ for a homogeneous one-dimensional problem.}
    \label{fig:model}
\end{figure}

Further insight into the model behaviour is provided here by considering the solution to the one-dimensional problem. Thus, a homogeneous material state is assumed, which means that the Laplacian term in the left side of \Cref{eq:localbalance_phi} vanishes. For illustrative purposes, the yield stress, hardening exponent, and fracture length scale are taken as $\sigma_{y0}=800$ MPa, $n=0.1$, and $\ell=0.4$ mm, respectively. In \Cref{fig:model}, the stress-strain response is plotted for various values of the critical energy release rate, $G_c$.\\

For a purely elastic material, the maximum obtained stress from this one-dimensional homogeneous solution can be expressed as \cite{Tanne2018,Martinez-Paneda2018}
\begin{equation}
    \sigma_c = \sqrt{\frac{27EG_c}{256\ell}}.
    \label{eq:sigma_c}
\end{equation}
The values of this elastic failure stress for the cases considered are 2054 MPa, 1703 MPa, 1258 MPa, and 513 MPa, corresponding to $G_c$ values of 80 kJ/m$^2$, 55 kJ/m$^2$, 35 kJ/m$^2$, and 5 kJ/m$^2$, respectively. For all but the last case, $\sigma_c$ is higher than the initial yield stress, $\sigma_{y0}$. In this situation, the material yields and the plastic dissipation starts to contribute to the damage process through the term $\psi_p$ in \Cref{eq:localbalance_phi}. The material fails when the combined contribution from the elastic and plastic energies approaches $G_c$. The maximum attained stresses for these three cases are close to each other. However, a decreased $G_c$ significantly decreases the elongation at failure. This effect of hydrogen on the stress-strain curve is in good agreement with what is typically observed in experiments.\\

For the case where $Gc=5$ kJ/m$^2$, the elastic failure stress of $\sigma_c=513$ MPa is lower than the yield strength of $\sigma_{y0}=800$ MPa. In this situation, the material fails at $\sigma = \sigma_c$ and no plastic deformation is predicted. The ductility ratio, $r_y=\sigma_c/\sigma_{y0}$, is commonly adopted to describe the degree of ductility of a material. $r_y>1$ indicates ductile behavior, while $r_y<1$ indicates brittle behavior.\\ 

Note that the observations made in this section are only strictly valid for the one-dimensional case, since yield and fracture are driven by different projections of the three-dimensional stress state. For example, some local yielding might still occur in materials where $r_y<1$. Nevertheless, $r_y$ remains a useful indicator of the ductility of a material.

\section{Transition flaw size analysis}
\label{section:flawsize}

The competition and interaction between toughness and strength driven failures become evident in a transition flaw analysis \cite{Usami1986,Chen2017}. For a uniform stress state, a material fails at a critical nominal stress $\sigma_u$, while for large pre-existing cracks, material failure is driven by crack propagation, which is governed by the energy release rate. For intermediate crack sizes, a mediation between the two limiting failure criteria occurs, with plasticity playing a role.\\

Tanne et al. \cite{Tanne2018} and Kristensen et al.~\cite{kristensen2021assessment} demonstrated that phase field fracture models capture both failure modes and the transition between them for elastically brittle materials. Here, we extend this analysis to ductile materials. The SENT specimen depicted in \Cref{fig:sent_sketch} is modeled with varying initial crack lengths, $a_0$, while the material properties are kept constant. The model parameters are chosen such that the ductility ratio has a value of $r_y=1.5$, representing a ductile material. Under a uniform stress state, the material fails due to plastic collapse at an ultimate stress, $\sigma_u$. In contrast to brittle materials, as discussed in \ref{section:appendix_model}, no analytical expression for $\sigma_u$ is available. Therefore, $\sigma_u$ is determined as the maximum attained stress in the one-dimensional stress-strain curve, of which examples are shown in \Cref{fig:model}.\\

\begin{figure}[!tb]
    \begin{subfigure}{\linewidth}
        \includegraphics[width=\linewidth]{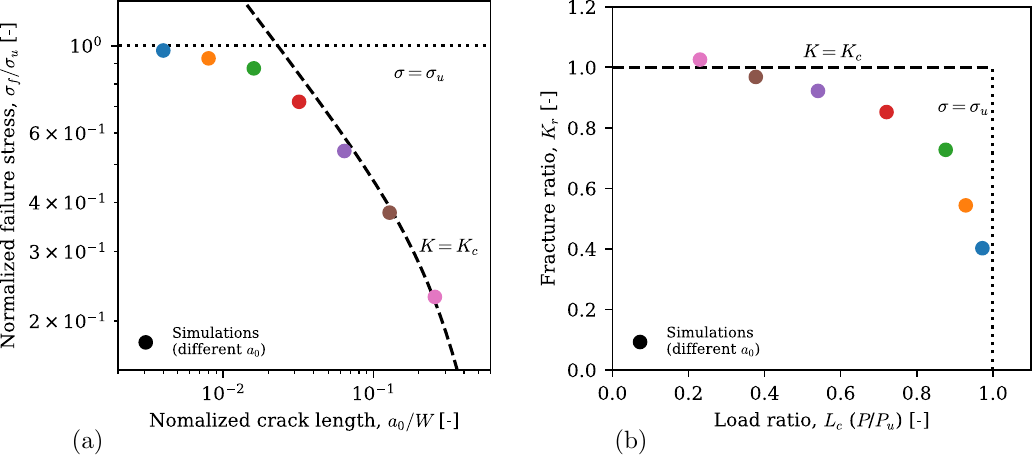}
        \phantomcaption \label{fig:flawsize}
        \phantomcaption \label{fig:flawsize_interaction}
    \end{subfigure}
    \caption{Transition flaw size analysis. (a) Failure stress for different crack lengths. (b) Interaction diagram for different crack lengths.}
    \label{fig:flawsize_full}
\end{figure}

\Cref{fig:flawsize} presents the predicted remote failure stresses, $\sigma_f$, normalized by $\sigma_c$, for the SENT specimen for different crack sizes. In addition, the elastic Griffith's criterion ($K=K_c$) and the plastic collapse criterion ($\sigma=\sigma_u$) are depicted in \Cref{fig:flawsize}. For short cracks, the values of the failure stress obtained are close to those associated with the plastic collapse criterion (strength-controlled failure). For longer initial cracks, a good agreement is obtained with Griffith's criterion for brittle materials. In between, a transition region is present where the material fails earlier than predicted by both failure criteria. In this region, failure is driven by an interaction between Griffith's criterion and plasticity.\\

The same data can be visualized in an interaction diagram, where the driving forces for both failure criteria are represented along an axis, which is shown in \Cref{fig:flawsize_interaction}. On the horizontal axis, the load ratio $L_c = P/P_u = \sigma / \sigma_u$ represents the proximity to plastic collapse, while on the vertical axis the fracture ratio $K_r=K/K_c$ represents the proximity to brittle crack propagation. The effects of plasticity on crack propagation become evident by considering the failure envelope of the simulation failure points. When even larger $a_0/W$ ratios than the ones in \Cref{fig:flawsize_full} are considered, failure occurs later than predicted by Griffith's theory. In these situations, the initial crack size is too large with respect to the geometry dimensions and edge effects dominate.

\end{document}